\documentclass{moriond}

\def\Journal#1#2#3#4{{#1} {\bf #2}, #3 (#4)}

\def\APJ{\em Astrophys. J.}
\def\CQG{\em Class. Quantum Grav.}
\def\AA{\em Astro. \& Astrophys.}

\def\be{\begin{equation}}
\def\ee{\end{equation}}
\def\bea{\begin{eqnarray}}
\def\eea{\end{eqnarray}}

\newcommand{\Photo}{}

\begin{document}
\vspace*{4cm}
\title{LOW LATENCY SEARCH FOR COMPACT BINARY COALESCENCES USING MBTA}

\author{T. ADAMS\\
on behalf of the LIGO Scientific Collaboration and the Virgo Collaboration}

\address{Laboratoire d'Annecy-le-Vieux de Physique des Particules (LAPP), Universit\'e de Savoie, CNRS/IN2P3, F-74941 Annecy-le-Vieux, France }

\maketitle\abstracts{The Multi-Band Template Analysis is a low-latency analysis pipeline for the detection of gravitational waves to triggering electromagnetic follow up observations. Coincident observation of gravitational waves and an electromagnetic counterpart will allow us to develop a complete picture of energetic astronomical events. We give an outline of the MBTA pipeline, as well as the procedure for distributing gravitational wave candidate events to our astronomical partners. We give some details of the recent work that has been done to improve the MBTA pipeline and are now making preparations for the advanced detector era.
}

\section{Introduction}

Currently the LIGO \cite{Abbott:2009li} and Virgo \cite{Accadia:2012vi} detectors are being brought back into operation after an extended period of upgrades and commissioning.
This is an extremely exciting time in the gravitational wave (GW) community as we prepare for the beginning of the advanced detector era, when advanced LIGO \cite{Aasi:2015al} and advanced Virgo \cite{Acernese:2015av} come online.
The advanced detectors will have a sensitive range for binary neutron stars (BNS) source a factor of $10$ better than the initial detectors, which corresponds to an increase in the observable volume by three orders of magnitude.
This will improve the ``realistic'' BNS detector rates from $0.02\,\textrm{yr}^{-1}$ in the initial detector era to $40\,\textrm{yr}^{-1}$ in the advanced detector era \cite{Abadie:2010fn}.

The first scheduled observing run in the advanced detector era will be a three month run starting in September 2015 which will only include the two LIGO detectors while the Virgo detector finishes its upgrades and commissioning.
Virgo will join the detector network for the first three detector observing run in 2016-2017 for six month.

\begin{figure}
  \centering
  \includegraphics[width=0.7\textwidth]{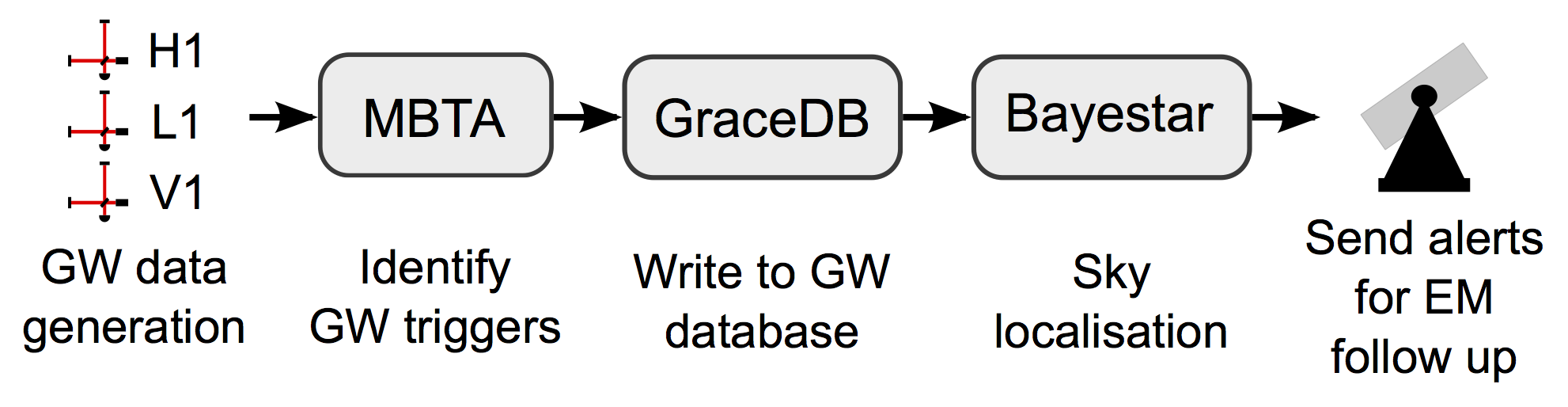}
  \caption{\label{fig:timeline}Overview of the joint GW-EM observation pipeline.}
\end{figure}

In this paper we will outline the key elements of the Multi-Band Template Analysis (MBTA) pipeline and the procedure for distributing GW candidate events for electromagnetic (EM) follow up observations, as shown in figure\,\ref{fig:timeline}.
In section\,\ref{sec:single_det} we explain how GW signals are extracted from the GW channel data of each detector.
In section\,\ref{sec:coincidence} we explain the criteria for selecting GW candidate events.
In section\,\ref{sec:followup} we give details of the post-MBTA event processing and explain how GW candidate events are distributed to our astronomical partners.
In section\,\ref{sec:improvements} we give details of some of the improvements recently implemented in MBTA.
In section\,\ref{sec:conclusion} we summarise the status of the MBTA pipeline in preparation for the advanced detector era.

\section{The Multi-Band Template Analysis}

MBTA \cite{Abadie:2012wx} is a low-latency coincidence analysis pipeline used to detect GWs from compact binary coalescences (CBCs).
Some of the best understood and strongest sources of GWs for the LIGO and Virgo detectors are CBCs consisting of two neutron stars (NS-NS), or a neutron star and a black hole (NS-BH).
These systems have many possible mechanisms for producing EM counter parts \cite{Metzger:2011wa}.
The possibility of a strong GW signal and EM counterpart make these sources the focus of the MBTA pipeline, which aims to detect GW candidate events with low enough latency to trigger EM follow up observations by our astronomical partners \cite{Aasi:2013lo}.

\subsection{Single detector analysis}
\label{sec:single_det}

Each detector in the network is analysed independently, before the results are later combined to find GW candidate events.
We obtain the calibrated GW channel data from each of the detectors, as well as basic data quality information, informing us of the status of the detectors.
MBTA uses the standard matched filter \cite{Waistein:1962vz} to extract CBC signals from the GW channel data of each detector.
To do this a bank of search templates is used to cover the parameter space of expected signals, this is generated at initialisation to keep the analysis latency as low as possible.
This template bank is referred to as the ``virtual'' template bank, and covers the parameter space we are interested in.

To reduce the computational cost of the matched filtering, which is the most computationally expensive element of the analysis, MBTA splits the matched filter across two (or more) frequency bands.
The boundary frequency between the low and high frequency bands is selected so that the signal-to-noise ratio (SNR) is roughly equally shared between the two bands.
This multi-band analysis procedure gives a reduction in the computational cost and we loose negligible SNR compared to a matched filter performed with a single band analysis.
The reduction in computational cost comes from the fact that in each frequency band we can use shorter templates and so the phase of the signal is tracked over fewer cycles, this reduces the number of templates that is required to cover the equivalent mass space of a single band analysis.
We also benefit from being able to use a reduced sampling rate for the low frequency band, which reduces the computational cost of the fast Fourier transforms involved in the filtering.

Each frequency bands requires a separate ``real'' template bank, which is actually used to filter the data.
For each template in the virtual template bank, a template from the low and high frequency real template banks are combined to reconstruct the result from a single band analysis.
The real template banks and the parameters for combination and association with the virtual template bank are produced during initialisation to keep the analysis latency low.

To further reduce the computational cost of the filtering, the template banks are split across multiple jobs and run in parallel across the parameter space.
Once we have the results from each band across the full parameter space the results are coherently combined between the bands.
Triggers are extracted from the match filter output in each band when $SNR > 5$, and a computationally inexpensive $\chi^{2}$ test is used to check the SNR distribution across the frequency bands is consistent with the expected signal SNR evolution.

\subsection{Coincidence events}
\label{sec:coincidence}

The trigger lists from the individual detectors are combined to find coincidence events.
Time coincidence it checked using a simple time of flight consistency test between triggers in detector pairs.
In the past we also used a mass coincidence criterion, but this has now been superseded by the exact match requirement.
The exact match requires that triggers are found in all detectors with the same template parameters; the component masses and spins.
The significance of each event is estimated by calculating the false alarm rate (FAR), the expected rate of coincidence triggers from noise only that have an equal or large SNR than the event.

\subsection{Event follow up}
\label{sec:followup}

\begin{figure}
  \centering
  \includegraphics[width=0.5\textwidth]{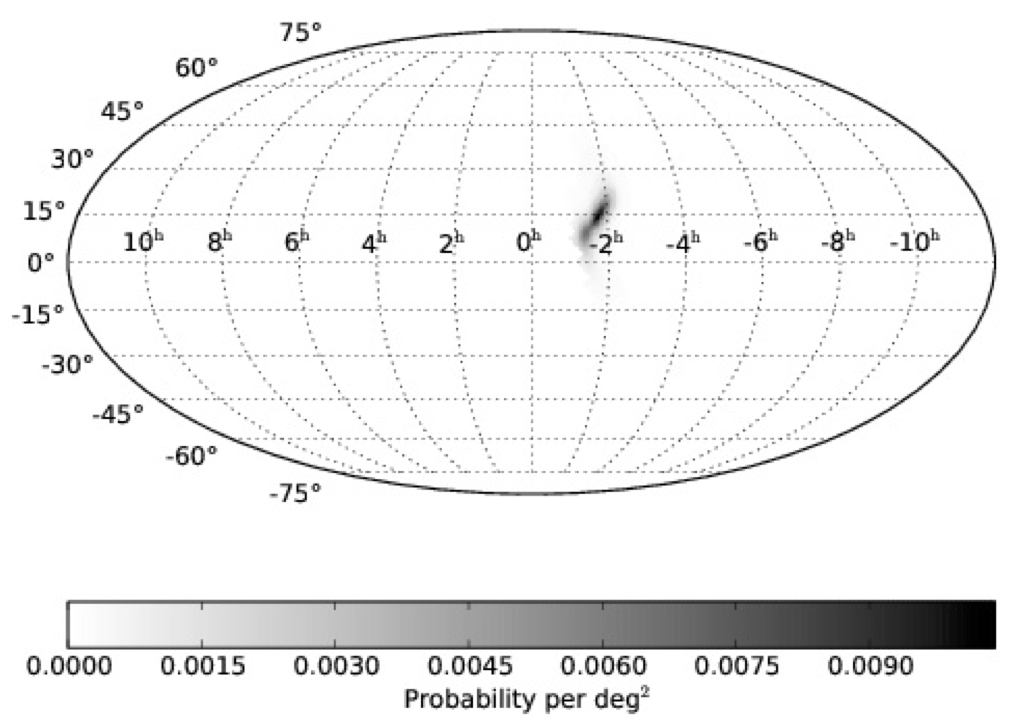}
  \caption{\label{fig:G20190}Skymap of GW candidate event G20190}
\end{figure}

GW candidate events found by the MBTA pipeline are sent to the Gravitational Wave Candidate Event Database (GraCEDb) \cite{gracedb}, an automated archive for GW candidate events.
When MBTA uploads a new event to GraCEDb rapid sky localisation is performed with Bayestar \cite{Singer:2014iq}, a rapid Bayesian position reconstruction code, using the time, amplitude, and phase information reported by MBTA.
Bayestar returns a probability skymap that is appended to the GraCEDb event and can be used to plan follow up observations around the most probably sky positions of the source.
As well as sky localisation, any additional data quality vetoes can be applied to remove events that are associated with detector noise.

In previous analysis runs \cite{Abadie:2012wx} GW candidate events have then undergone verification by a human monitor.
The purpose of this human monitor was to review the events, consult the detector control rooms about the status of the detectors, and verify the data quality.
In the past the entire pipeline, from data collection distributing GW candidate events to our astronomical partners, had a latency of 20-40 minutes; where the human monitor step gave the largest contribution to the latency.
We give an example of a GW candidate event skymap \cite{Abadie:2012wx} in figure\,\ref{fig:G20190}.
This event was distributed to our astronomical partners with a latency of 39 minutes and follow up observations were performed by Quest, ROTSE, SkyMapper, TAROT and Zadko.

\subsection{Recent improvements}
\label{sec:improvements}

In preparation for the first observing run in the advanced detector era we have implemented a number of improvements to the MBTA pipeline.
We are now able to run MBTA using spinning template banks, this was achieved by changing the interface for generating the template banks.
Now that we are using the exact match, we can provide the coalescence phase of events for improved sky localisation with Bayestar.
Finally, to remove noise events we have added a signal based consistency test.
This uses the fact that a real CBC signal should produce a single loud peak in the matched filter output, whereas a noise trigger could possibly produce multiple loud peaks.
Comparing the peak SNR of an event to the surrounding level we can veto events that do not behave like real CBC signals.

\section{Conclusion}
\label{sec:conclusion}

In this paper we have outlined the MBTA pipeline, and how the GW candidate events produced by MBTA are distributed to our astronomical partners for EM follow up observations.
As was shown in previous observing runs, we are ready to perform this task and are now focusing on implementing improvements in preparation for the first advanced era observing run in September 2015.
We expect that as the detectors sensitivities are improved, we will soon be making the first GW detections and can begin to do real GW astronomy in coalition with our astronomical partners.

\section*{Acknowledgements}

The authors gratefully acknowledge the support of the United States National Science Foundation (NSF) for the construction and operation of the LIGO Laboratory, the Science and Technology Facilities Council (STFC) of the United Kingdom, the Max-Planck-Society (MPS), and the State of Niedersachsen/Germany for support of the construction and operation of the GEO600 detector, the Italian Istituto Nazionale di Fisica Nucleare (INFN) and the French Centre National de la Recherche Scientifique (CNRS) for the construction and operation of the Virgo detector. The authors also gratefully acknowledge research support from these agencies as well as by the Australian Research Council, the International Science Linkages program of the Commonwealth of Australia, the Council of Scientific and Industrial Research of India, Department of Science and Technology, India, Science \& Engineering Research Board (SERB), India, Ministry of Human Resource Development, India, the Spanish Ministerio de Economia y Competitividad, the Conselleria dEconomia i Competitivitat and Conselleria dEducaci, Cultura i Universitats of the Govern de les Illes Balears, the Foundation for Fundamental Research on Matter supported by the Netherlands Organisation for Scientific Research, the Polish Ministry of Science and Higher Education, the FOCUS Programme of Foundation for Polish Science, the European Union, the Royal Society, the Scottish Funding Council, the Scottish Universities Physics Alliance, the National Aeronautics and Space Administration, the Hungarian Scientific Research Fund (OTKA), the Lyon Institute of Origins (LIO), the National Research Foundation of Korea, Industry Canada and the Province of Ontario through the Ministry of Economic Development and Innovation, the National Science and Engineering Research Council Canada, the Brazilian Ministry of Science, Technology, and Innovation, the Carnegie Trust, the Leverhulme Trust, the David and Lucile Packard Foundation, the Research Corporation, and the Alfred P. Sloan Foundation. The authors gratefully acknowledge the support of the NSF, STFC, MPS, INFN, CNRS and the State of Niedersachsen/Germany for provision of computational resources. This research has made use of the SIMBAD database, operated at CDS, Strasbourg, France.

\end{document}